# Digitized adiabatic quantum computing with a superconducting circuit

R. Barends,[1] A. Shabani,[2] L. Lamata,[3] J. Kelly,[1] A. Mezzacapo,[3, *] U. Las Heras,[3] R. Babbush,[2] A. G. Fowler,[1] B. Campbell,[4] Yu Chen,[1] Z. Chen,[4] B. Chiaro,[4] A. Dunsworth,[4] E. Jeffrey,[1] E. Lucero,[1] A. Megrant,[4] J. Y. Mutus,[1] M. Neeley,[1] C. Neill,[4] P. J. J. O'Malley,[4] C. Quintana,[4] P. Roushan,[1] D. Sank,[1] A. Vainsencher,[4] J. Wenner,[4] T. C. White,[4] E. Solano,[3,5] H. Neven,[2] and John M. Martinis[1,4]

[1] Google Inc., Santa Barbara, CA 93117, USA
[2] Google Inc., Venice, CA 90291, USA
[3] Department of Physical Chemistry, University of the Basque Country UPV/EHU, Apartado 644, E-48080 Bilbao, Spain
[4] Department of Physics, University of California, Santa Barbara, CA 93106, USA
[5] IKERBASQUE, Basque Foundation for Science, Maria Diaz de Haro 3, 48013 Bilbao, Spain.

A major challenge in quantum computing is to solve general problems with limited physical hardware. Here, we implement digitized adiabatic quantum computing, combining the generality of the adiabatic algorithm with the universality of the digital approach, using a superconducting circuit with nine qubits. We probe the adiabatic evolutions, and quantify the success of the algorithm for random spin problems. We find that the system can approximate the solutions to both frustrated Ising problems and problems with more complex interactions, with a performance that is comparable. The presented approach is compatible with small-scale systems as well as future error-corrected quantum computers.

Quantum mechanics can help solve complex problems in physics [1], chemistry [2], and machine learning [3], provided they can be programmed in a physical device. In adiabatic quantum computing [4–6], the system is slowly evolved from the ground state of a simple initial Hamiltonian to a final Hamiltonian that encodes a computational problem. The appeal of this analog method lies in its combination of simplicity and generality; in principle, any problem can be encoded. In practice, applications are restricted by limited connectivity, available interactions, and noise. A complementary approach is digital quantum computing, where logic gates combine to form quantum circuit algorithms [7]. The digital approach allows for constructing arbitrary interactions and is compatible with error correction [8, 9], but requires devising tailor-made algorithms. Here, we combine the advantages of both approaches by implementing digitized adiabatic quantum computing in a superconducting system. We tomographically probe the system during the digitized evolution, explore the scaling of errors with system size, and measure the influence of local fields. We conclude by having the full system find the solution to random Ising problems with frustration, and problems with more complex interactions. This digital quantum simulation [10–13] consists of up to nine qubits and up to $10^3$ quantum logic gates. This demonstration of digitized quantum adiabatic computing in the solid state opens a path to solving complex problems, and we hope it will motivate further research into the efficient synthesis of adiabatic algorithms, on small-scale systems with noise as well as future large-scale quantum computers with error correction.

A key challenge in adiabatic quantum computing is to construct a device that is capable of encoding problem Hamiltonians that are non-stoquastic [14]. Such Hamiltonians would allow for universal adiabatic quantum computing [15, 16] as well as improving the performance for difficult instances of classical optimization problems [17]. Additionally, simulating interacting fermions for physics and chemistry requires non-stoquastic Hamiltonians [1, 18]. In general, non-stoquastic Hamiltonians are more difficult to study classically, as Monte Carlo simulations fail to converge due to the "sign problem" [19]. A hallmark of non-stoquastic Hamiltonians is the need for several distinct types of coupling, for example containing both $\sigma_z\sigma_z$ and $\sigma_x\sigma_x$ couplings with different signs. With a digitized approach different couplings can be constructed without change of hardware. In addition, long-range multibody interactions can be assembled to aid in quantum tunneling [20] or to encode the non-local terms required for fermionic simulations [21, 22], which is otherwise hampered by limited connectivity. And finally, analog systems exhibit noise which can thwart the evolution, whereas digital systems can have error correction. Our experiment addresses the open challenge of adiabatically evolving to final problem Hamiltonians that are non-stoquastic.

Here, we explore the adiabatic quantum evolutions of one-dimensional spin chains with nearest-neighbour coupling. We start with a simple ferromagnetic problem to visualize the adiabatic evolution process. We identify specific error contributions, and follow up by exploring the scaling of errors with system size. We move towards more complex Hamiltonians and add local fields. We finish by testing the device on random problems that are stoquastic, containing $\sigma_z\sigma_z$ type coupling, as well as non-stoquastic problems with both $\sigma_z\sigma_z$ and $\sigma_x\sigma_x$ couplings. The initial and problem Hamiltonians are

$$H_I = -B_{x,I}\sum_i \sigma_x^i, \tag{1}$$

$$H_P = -\sum_i \left(B_z^i\sigma_z^i + B_x^i\sigma_x^i\right) \tag{2}$$
$$ -\sum_i \left(J_{zz}^{i,i+1}\sigma_z^i\sigma_z^{i+1} + J_{xx}^{i,i+1}\sigma_x^i\sigma_x^{i+1}\right),$$

where $B_z$ and $B_x$ denote local field strengths, and $J_{zz}$ and $J_{xx}$ the $\sigma_z\sigma_z$ and $\sigma_x\sigma_x$ coupling strengths. The Ising model is recovered when $B_x = J_{xx} = 0$. We initialize the system with Hamiltonian $H_I$ whose ground state is trivial to prepare (for $B_{x,I} = 2$ the ground state is $|+\rangle^{\otimes N}$) and vary the system Hamiltonian to the final problem Hamiltonian: $H = sH_P + (1-s)H_I$, with $s$ going from 0 to 1. An example problem is shown in Fig. 1a.

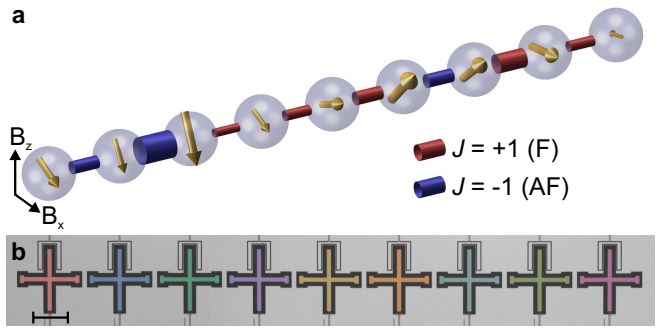

FIG. 1. **Spin chain problem and device.** (a) We implement one-dimensional spin problems with variable local fields and couplings between adjacent spins. Shown is an example of a stoquastic problem Hamiltonian with local $X$ and $Z$ fields, indicated by the gold arrows in the spheres, and $\sigma_z \sigma_z$ couplings, whose strength is indicated by the radius of the links. Red denotes a ferromagnetic and blue an antiferromagnetic link. Problem is for the instance in Fig. 5c. (b) Optical picture of the superconducting quantum device with nine Xmon qubits $Q_0$ - $Q_8$ (false-coloured cross-shaped structures), made from Al (light) on a sapphire substrate (dark). Connections to readout resonators are at the top, and control wiring is at the bottom. Scale bar is $200~\mu m$.

The spin system is formed by a superconducting circuit with nine qubits. The qubits are the cross-shaped structures [23], patterned out of an Al layer on top of a sapphire substrate, and arranged in a linear chain, see Fig. 1b. Each qubit is capacitively coupled to its nearest neighbours, and can be individually controlled and measured; for details see Ref. [24]. Crucially, by tuning the frequencies of the qubits we can implement a tunable controlled-phase entangling gate, which together with the single qubit gates forms our digitized approach. We use the Trotter expansion to digitize [25], the evolution is divided into many steps in time, each of which is implemented using a construction of quantum gates, see Supplementary Information.

For quantifying digitized adiabatic evolutions there are four sets of data: I) The ideal continuous time evolution, for infinite time, which is free of error and provides the perfect solution; we refer to this as "target state". II) The ideal continuous time evolution for a finite time $T$, which is sensitive to non-adiabatic errors. We call these results: "ideal continuous evolution". III) The "ideal digital evolution", where the finite ideal continuous evolution is digitized, and therefore includes digital error as well as non-adiabatic errors. And IV) the experimental results, which include a contribution from gate errors as well.

We start with a ferromagnetic chain problem with $N = 4$ spins, and equal coupling strength $J_{zz} = 2$. The qubits are initialized in the $|+\rangle^{\otimes N}$ state, and we use five steps to evolve the system to the problem Hamiltonian, performing quantum state tomography after each step. We linearly decrease the $B_x$ term to zero, starting at $B_x = 2$, and simultaneously increase the coupling strength from 0 to 2, ending the evolution at a scaled time of $|J|T = 6$. The density matrices are shown in Fig. 2a. With each step the quantum state evolves and matrix elements in the middle vanish while the elements at the four corners grow to form the density matrix $\rho$ of the Greenberger-Horne-Zeilinger (GHZ) state a fidelity $\mathrm{Tr}(\rho_{\mathrm{target-state}}\rho)$ of 0.55. The density matrix is constrained to be physical [26]. The ideal digital evolution is plotted in Fig. 2b, reaching a fidelity of 0.85. The Hamiltonian during evolution, construction of the algorithm, and pulse sequence are shown in Figs. 2c-e. In each Trotter step, we perform a $\sigma_z \sigma_z$ operation on each pair, to implement the ferromagnetic $\sigma_z \sigma_z$ coupling, followed by single qubit rotations around the $X$ axis to simulate the transversal magnetic field. In the pulse sequence, the rectangular-like frequency detuning pulses indicate where $\sigma_z \sigma_z$ interaction is implemented by bringing qubits near resonance (highlighted for $s = 0.2$). The wave-like pulses are microwave gates; the decrease in $B_x$ with Trotter steps is reflected by the reduction in amplitude of the corresponding microwave pulses (highlighted for $s = 0.4$ and $s = 1.0$). Additional microwave echo pulses decrease coupling to other qubits and the environment. We find mean phase errors from neighbouring parasitic interactions to be around 0.05 rad, equivalent to an error contribution below $10^{-3}$ (see Supplementary Information).

The experiment in Fig. 2 shows that digital synthesis of adiabatic evolutions can successfully be implemented in a solid state quantum platform. Using five Trotter steps, 15 entangling gates and 144 single-qubit microwave gates, the system forms diagonal as well as off-diagonal terms which are close to the ideal, finding a genuinely entangled GHZ state, the exact solution to the ferromagnetic problem. It shows that complex pulse sequences are possible, and that the errors make sense. The fidelity of the experimental data with respect to the ideal digital evolution is 0.64. The fidelity of the ideal digital evolution with respect to the ideal continuous time evolution (equivalent to an infinite number of Trotter steps) has a fidelity 0.93, and the overlap of the continuous evolution with the GHZ state (see Supplementary Information) is 0.88. The product of the above three values (0.52) is close to the experimental fidelity of 0.55, and shows the experimental error is a combination of non-adiabatic errors, digitization errors, as well as gate errors. Adopting the entangling gate error of $7.4 \cdot 10^{-3}$ and $8 \cdot 10^{-4}$ as measured in Ref. [26], we expect an accumulated gate error of 0.23 whereas we find an infidelity of 0.36; we attribute the difference to errors in maintaining the phases of the four qubit system for a duration of 2.1 $\mu s$.

An important feature of the errors is the prevalence of populations and correlations of the $|0001\rangle$, $|0011\rangle$, and $|0111\rangle$-states and their bitwise inverse, see arrows in Fig. 2a. Their elements are also present in the ideal digital results as well as the ideal continuous evolutions (see Supplementary Information). These are states that deviate by a single kink from the target state, having a residual energy of $2|J|$, indicating the presence of non-adiabatic errors. These kink errors are connected to the formation of defects during a phase transition, as

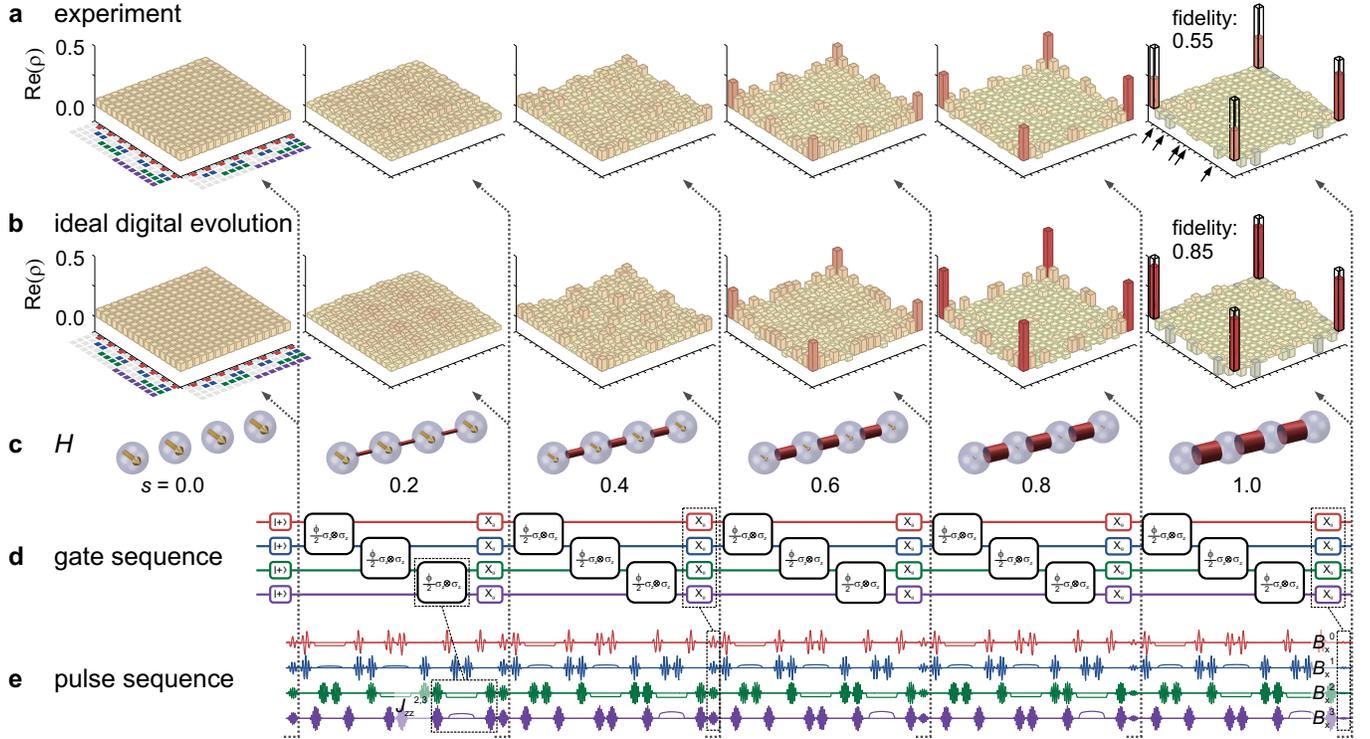

FIG. 2. **Quantum state tomography of the digital evolution into a Greenberger-Horne-Zeilinger state.** A four qubit system is adiabatically evolved from an initial Hamiltonian where all spins are aligned along the $X$ axis to a problem Hamiltonian with equal ferromagnetic couplings between adjacent qubits ($J_{zz} = 2$). (a) Experimental density matrix $\rho$ at the start and after each Trotter step, showing the growth of the major elements on the four corners, measured using quantum state tomography. The target state is shown in black. Coloured squares indicate qubit indices: For example, $Q_0$ being excited is indicated by a red square. Black arrows indicate significant elements for states which differ from the target state by a single kink. (b) Ideal digitized evolution, showing major elements on the four corners as well as other populations and correlations. Real parts shown. (c) Hamiltonian at different $s$. (d) Gate sequence, showing initialization and the five Trotter steps. (e) Pulse sequence, showing the single-qubit microwave gates (wave-like pulses) and frequency detuning pulses (rectangular-like). Corresponding interactions and local field terms are highlighted. The displayed five step algorithm is 2.1 $\mu s$ long. Colours correspond to the physical qubits in Fig. 1b. Implementations of $\sigma_z\sigma_z$ coupling and local $X$-fields are highlighted. See Supplementary Information for imaginary parts and the ideal continuous evolution.

described by the Kibble-Zurek mechanism [27, 28].

To explore the scaling of errors we vary the system size from two to nine qubits and measure the likelihood of kinks and residual energy. We keep the ferromagnetic problem Hamiltonian, $J_{zz} = 2$, but vary the scaled time such that $|J|T$ goes from 0 to 3. For the two to six qubit system we use five Trotter steps and for seven to nine qubits we use two steps, to limit the total number of gates. The kink likelihood for the four qubit system is shown in Fig. 3a. Here, the likelihood of one kink is given by the sum of the probabilities of all states with one kink (e.g. $|0001\rangle$, $|0011\rangle$, and $|0111\rangle$ and their bitwise inverse). When increasing $|J|T$ from 0 to 3 the kink likelihood decreases, and the likelihood of no kinks increases (black line). The experimental data closely follow the ideal digital evolution (dashed). This picture is repeated for all systems, see Supplementary Information.

The kink likelihood signals that the final state has residual energy, as a state with a single kink has energy $2|J|$ above the target state. The residual energies for all systems are plotted in Fig. 3b. Initially, the residual energy is constant at $|J|T \sim 0$, and starts to decrease around $|J|T \sim 0.5$, follow-ing both the ideal digital (dashed) and ideal continuous time evolutions (dotted). For two to six qubits, this decrease continues until the traces start to settle around $|J|T = 3$. For the seven to nine qubit system, the residual energy starts to increase again around $|J|T = 2$, following the ideal digital evolution. See Supplementary Information for the pulse sequence for the nine qubit experiment, all kink likelihoods, and for the differences between the residual energies.

The main result is that Fig. 3 distinctly shows the different contributions to error (highlighted): For $|J|T \ll 1$, the residual energy is dominated by non-adiabatic errors as the evolution moves too fast. For $|J|T > 2$, the flattening out of the residual energy for the configurations with two to six qubits indicates that gate errors dominate, as the predictions from the ideal digital evolutions are significantly lower. And for the larger qubit configurations with seven to nine qubits, the residual energy follows the digital predictions upwards, indicating that digitization errors dominate. In addition, the residual energy visibly decreases at $|J|T = 1$ for all configurations, implying that the digitized evolutions are able to approach the target state even for nine qubits.

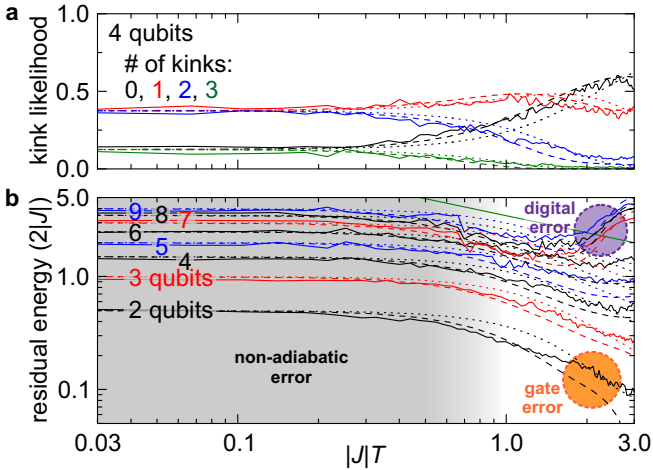

FIG. 3. **Kink errors, residual energy, and scaling with system size.** (a) Kink likelihood for the four qubit configuration. Solid lines: experiment. Dashed lines: ideal digital evolution. Dotted lines: ideal continuous time evolution. (b) Residual energy in the adiabatic evolutions of ferromagnetic chains ($J_{zz} = 2$) in configurations with two to nine qubits. The green solid line shows the ideal square-root trend for the large-scale limit (Supplementary Information). Distinct contributions to error are highlighted.

With the full system able to handle degenerate problem Hamiltonians, we now move towards more complex Hamiltonians and explore the lifting of degeneracy with the long-range effects of local fields. We apply a local $Z$ field ($B_Z$) on the middle qubit of a five-qubit antiferromagnetic chain. In Fig. 4a we show the single spin magnetization $< \sigma_z^i >$ for all qubits of index $i$ as a function of $B_z$. In the absence of a field, the state is degenerate, and the single qubit magnetizations are zero. With local field, the magnetization develops and displays the hallmark antiparallel configuration, following the ideal digital predictions (right). The experimental results are more pronounced for $B_z > 0$, we attribute this to gate errors arising from the asymmetry of implementation with sign, see Supplementary Information. We also plot in Fig. 4b the parity $< \sigma_z^i \sigma_z^{i+d} >$ with distance $d$, averaged over $B_z$. The mean correlation alternates sign (not shown) and decreases with distance, following the ideal trend. See Supplementary Information for the field-dependence of the parity.

The experiment in Fig. 4 shows that long-range correlations are generated in the system, even though the physical coupling of the system is nearest-neighbour only, and become visible when we apply local fields and lift the degeneracy of the antiferromagnetic state, creating classical Néel states.

We next discuss how the digitized approach can solve both stoquastic and non-stoquastic problems with comparable performance, by testing random problems on three, six, seven, eight and nine qubits. Problems have local fields and couplings with random strength and sign. We independently choose $B_z$ and $B_x$ from [-2, 2] for each spin, and $J_{zz}$ from [-2, -0.5] or [0.5, 2] for each link. This creates a random Ising problem with frustration. For non-stoquastic problems we also add $J_{xx}$ coupling for each link, with values from [-2, -0.5] or [0.5, 2]; effectively doubling the amount of entangling gates. We avoid small couplings to reduce the number of gates. For the three qubit systems we have used quantum state tomography on 100 separate instances, to include off-diagonal elements in the fidelity metrics. For six or more qubits, tomography is not practical and we have measured the correlated probabilities on 250 separate instances, and use a measure of success that sets an upper bound on the fidelity [29]. In Fig. 5 we show the results for stoquastic problems with three, six and nine spins, and non-stoquastic problems with three, six and seven spins. Per case, we highlight a single instance and show histograms of the fidelities.

For the three-spin stoquastic problems, the real part of the density matrix of one instance and a histogram of its diagonal elements are shown in Fig. 5a. In the tomography plot we overlay the experimental results (colour) with the ideal digital (black), and ideal continuous results (gray). For this example, we find fidelities $\mathrm{Tr}(\rho_{\mathrm{ideal-digital}}\rho) = 0.70$ and $\mathrm{Tr}(\rho_{\mathrm{target-state}}\rho) = 0.63$. In the top right, we show in colour the histograms for all instances of the fidelities $\mathrm{Tr}(\rho_{\mathrm{target-state}}\rho)$. Shown in gray is the fidelity of the ideal digital evolution with respect to the target state. Stoquastic problems with six and nine qubits are displayed in Figs. 5b-c. The main figures show the measured probabilities (colour) sorted by the target state results (gray), and the insets display the histograms. Non-stoquastic problem results are displayed on the right in Figs. 5d-f.

The key result from Figs. 5 is that the physical system can find solutions to non-stoquastic problems with a performance comparable to that of stoquastic problems. The three qubit examples show major diagonal as well as off-diagonal elements close to the expected positions. And visibly, for six and more qubit systems the coloured bars in the example instances are mostly on the left, indicating that the system has a clear preference for returning the probabilities associated with the ideal solutions.

The physical system produces results which are comparable to the expectations, as the histograms show a significant overlap between experiment and theory. Moreover, the num-bers are consistent, as we now discuss the six qubit stoquastic example. The mean success rate between the ideal adiabatic evolution and target state is $0.59 \pm 0.01$, indicating that the scaled time is large enough for capturing the evolution dynamics. The mean success rate of the ideal digitized evolution with respect to the ideal adiabatic evolution is $0.73 \pm 0.01$, indicating a proper Trotterization of the evolution. Finally, the value for the experimental evolution with respect to the ideal digitized evolution is $0.714 \pm 0.006$, indicating that the experiment follows the ideal digital evolution reasonably well. Interestingly the product of the above three numbers, $0.31$, is very close to the mean value between the experimental data and the target state, $0.296 \pm 0.007$. This shows that the experimental errors arise from comparable contributions of non-adiabatic, digital, and gate errors. For the six qubit non-stoquastic case, experimental-to-target state values are higher then this product, suggesting that errors partially cancel. A further reason for the measures of success being higher for non-stoquastic problems may be that the presence of $\sigma_x\sigma_x$ terms during adiabatic evolutions is helpful for difficult problems in general [17]. See Supplementary Information for a complete listing of fidelities and measures of success.

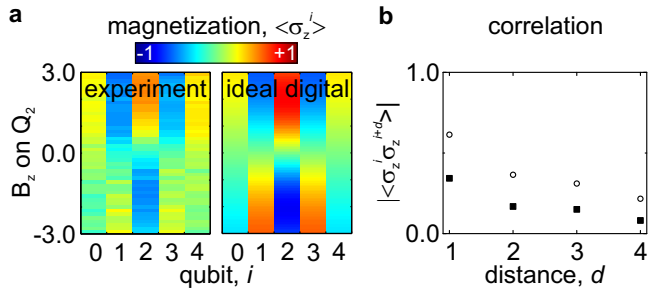

FIG. 4. **Lifting degeneracy of a five-spin antiferromagnetic state.** Antiferromagnetic ($J_{zz} = -2$) problem with a tunable local field on qubit 3. (a) Single qubit magnetization $< \sigma_z >$ as a function of magnetic field for the experiment (left), and ideal digital evolution (right). (b) Mean experimental (closed) and theory (open) spin parity correlation versus distance. The absolute value shows the long-range correlations.

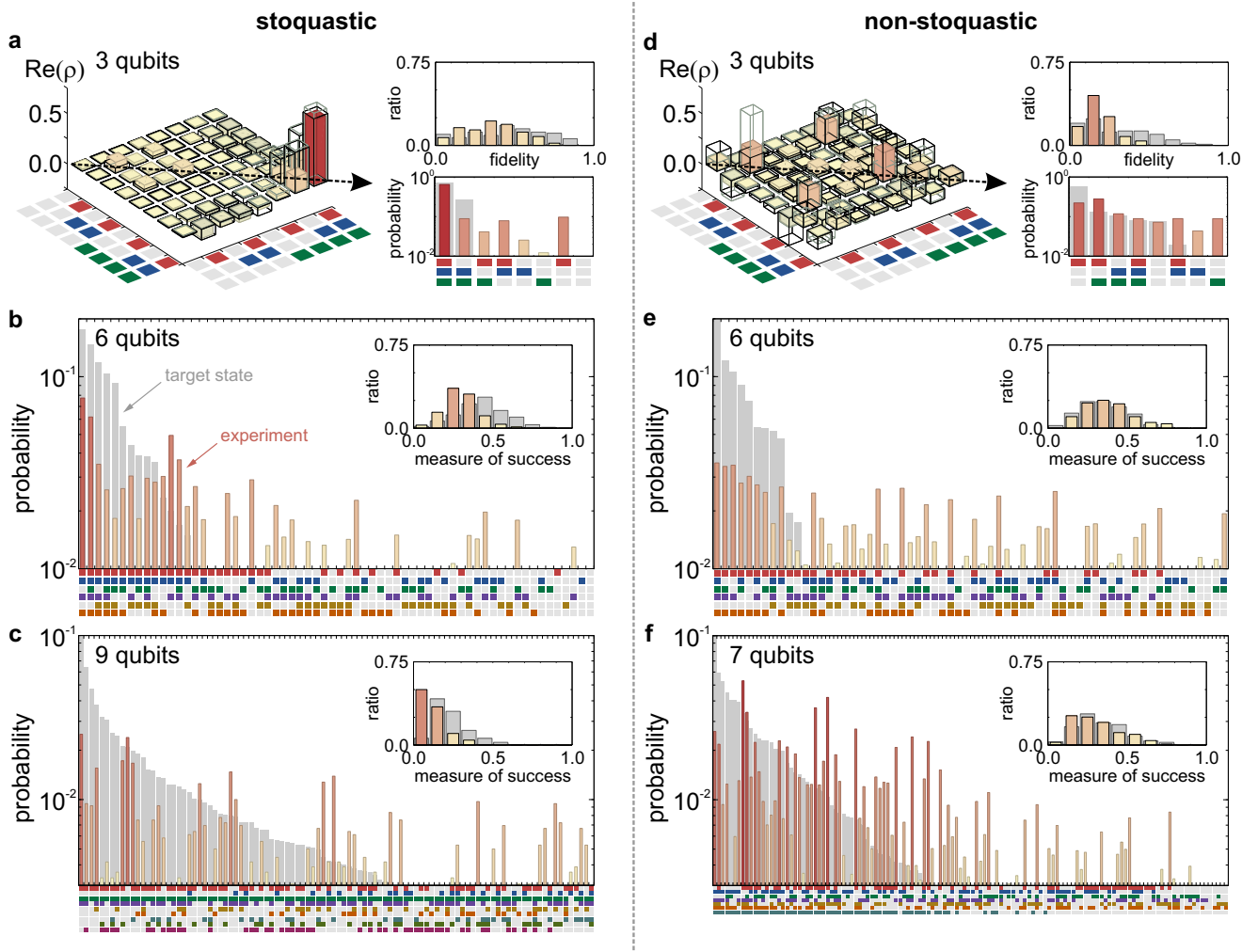

FIG. 5. **Digital evolutions of random stoquastic and non-stoquastic problems.** As stoquastic problems we use frustrated Ising Hamiltonians, having random local $X$ and $Z$ fields, and random $\sigma_z\sigma_z$ couplings. (a, b, c) Stoquastic results are shown for three, six and nine qubits. For three qubits we have done tomography. An example instance is on the left in (a), where we show the real part of the density matrix. Coloured bars denote the experimental data, black the ideal digital evolution, and gray the target state. The diagonals of the experiment (colour) and the target state (gray) are on the right, rank ordered by ideal target state results. The fidelity results for all 100 instances are summarized in the normalized histogram, top right. Coloured bars: fidelities of experimental results with respect to the target state. Gray: fidelities of the ideal digital evolution with respect to the target state. The correlated probabilities for six (b) and nine (c) qubits are plotted in the main figure, sorted by target state results. Experimental data is in colour, the target state is in gray. The results for all 250 instances are summarized in the insets. For the nine qubit instance the first 100 elements are shown. (d, e, f) Non-stoquastic problems have additional random $\sigma_x\sigma_x$ couplings, here we plot the data for three, six, and seven qubits. The results show that the system can find the ground states of both stoquastic and non-stoquastic Hamiltonians with comparable performance.

Finally, we note that this experiment took up to nine qubits and up to $10^3$ single and entangling quantum logic gates. The pulse sequences and gate counts, as well as spin problem parameters are in the Supplementary Information.

To further quantify the performance of the system we compare experimental as well as random probabilities with the theoretical results. In essence, we take a uniform random distribution as a baseline sanity check. We find that for the stoquastic problems, the fidelity of all six to nine qubit configurations are significantly above this baseline: for six qubits, the fidelity of the experimental data with respect to the target state is $0.296 \pm 0.007$, while using uniform random probabilities produces a value of $0.168 \pm 0.005$. For the nine qubit case the numbers are: $0.122 \pm 0.006$ for the experimental data and $0.074 \pm 0.004$ for random. For the non-stoquastic problems, the six and seven qubit configurations are above the baseline. A complete listing for all configurations can be found in the Supplementary Information.

This experiment shows that digital synthesis of the adiabatic evolutions can be used to find signatures of the ground states of random stoquastic as well as non-stoquastic problems. Errors arise from a comparable contribution of non-adiabatic, digital, and gate errors; and success rates are significantly above a uniform random baseline. We note that for larger qubit systems the number of Trotter steps needs to be limited to reduce the accumulation of gate error; in turn limiting the evolution we can simulate. The experimental error is therefore larger, from a combination of gate, digitization, and non-adiabatic error. However, in an error-corrected system the number of gates is in principle unconstrained, digitization can be made arbitrarily accurate, and one can move slower through critical parts of the evolution. While we have used Trotterization [30, 31], with recent methods based on the truncation of Taylor series [32] the scaling of the digitization becomes appealing. See Supplementary Information for further motivations and discussions.

In conclusion, we have experimentally demonstrated digitized adiabatic quantum evolutions in superconducting circuits: We tomographically verify the adiabatic evolution, and probe the effects of local fields. In addition, we show how the system can perform primitive problem solving, as it approximates the ground states of stoquastic as well as non-stoquastic problems. We believe that the digitized approach to adiabatic quantum evolutions of complex problems, where local fields, variable coupling strengths and types, as well as multibody interactions can be constructed, becomes viable on the small scale with lower gate errors, and that large scale applications can be done in conjunction with error correction. We hope our work accelerates improving superconducting quantum systems, and motivates further research into the encoding of and measurement for non-stoquastic computational problems. In addition, we hope these results encourage work on the efficient digitization of algorithms for small and large scale systems, where either reducing the effects of noise by for example dynamical decoupling techniques, or reducing the amount of T gates is of paramount importance.

**Acknowledgements** The authors acknowledge support from Spanish MINECO FIS2012-36673-C03-02; Ramón y Cajal Grant RYC-2012-11391; UPV/EHU UFI 11/55 and EHUA14/04; Basque Government IT472-10; a UPV/EHU PhD grant; PROMISCE and SCALEQIT EU projects. Devices were made at the UC Santa Barbara Nanofabrication Facility, a part of the NSF-funded National Nanotechnology Infrastructure Network, and at the NanoStructures Cleanroom Facility.

---

# Supplementary Information for "Digitized adiabatic quantum computing with a superconducting circuit"

R. Barends,[1] A. Shabani,[2] L. Lamata,[3] J. Kelly,[1] A. Mezzacapo,[3, *] U. Las Heras,[3] R. Babbush,[2] A. G. Fowler,[1] B. Campbell,[4] Yu Chen,[1] Z. Chen,[4] B. Chiaro,[4] A. Dunsworth,[4] E. Jeffrey,[1] E. Lucero,[1] A. Megrant,[4] J. Y. Mutus,[1] M. Neeley,[1] C. Neill,[4] P. J. J. O'Malley,[4] C. Quintana,[4] P. Roushan,[1] D. Sank,[1] A. Vainsencher,[4] J. Wenner,[4] T. C. White,[4] E. Solano,[3, 5] H. Neven,[2] and John M. Martinis[1, 4]

[1]*Google Inc., Santa Barbara, CA 93117, USA*
[2]*Google Inc., Venice, CA 90291, USA*
[3]*Department of Physical Chemistry, University of the Basque Country UPV/EHU, Apartado 644, E-48080 Bilbao, Spain*
[4]*Department of Physics, University of California, Santa Barbara, CA 93106, USA*
[5]*IKERBASQUE, Basque Foundation for Science, Maria Diaz de Haro 3, 48013 Bilbao, Spain.*

## CONTENTS

* Present address: IBM T. J. Watson Research Center, Yorktown Heights, NY 10598, USA

# I. WHY DIGITIZED ADIABATIC QUANTUM COMPUTING?

Implementing an adiabatic quantum algorithm on a gate-based quantum computer has been discussed in the original works introducing adiabatic quantum computing (AQC) [1, 2]. However, the motivation for those works was to investigate the power of AQC by mapping the adiabatic algorithm to the gate model [3–5]. In this work, we promote digitized AQC as a viable quantum algorithm for execution on an error corrected digital quantum device.

In contrast to conventional quantum algorithms, such as Shor's algorithm and Grover's algorithm [6], AQC is a general-purpose optimization algorithm. Any optimization problem, in principle, can be mapped and solved via AQC. However, AQC is a heuristic algorithm as there is no guarantee on the computation run time and it strongly depends on the nature of the problem. It is an active area of research to search for computational problems for which AQC yields a speed-up over its classical counterparts [7–12]. AQC is an approach to quantum computing that uses continuous dynamics. Therefore, building an analog processor to implement an adiabatic quantum evolution is a natural choice. Such a processor is commonly known as a quantum annealer. An analog quantum annealer has certain limitations that we propose can be overcome by a gate-model realization of the AQC. Here we list some key features that boosts the algorithmic success of a quantum annealer:

*Graph connectivity and k-body interactions-* These factors yield a computational landscape with tall and narrow energy barriers that make it easier for AQC to outperform those algorithms that use classical dynamics, such as simulated thermal annealing [13], spin Monte Carlo [14], and cluster finding based algorithms [15].

*Arbitrary interactions-* AQC finds the power of a universal quantum computer when it has programmable non-stoquastic Hamiltonian terms [3–5, 16]. A Hamiltonian is non-stoquastic when there is no representation in a standard basis with all non-positive off-diagonal terms [17, 18]. In the context of many-body physics, fermionic systems [19] or spin systems with Heisenberg XYZ interactions [18] are some examples for non-stoquastic Hamiltonians which suffer from the sign problem. Realizing arbitrary off-diagonal interactions is a signif-icant problem for analog systems as it requires perturbative gadgets with great precision [16, 20, 21].

*Precision-* Encoding computational problems in a quantum annealer such as the number partitioning problem [8, 22], requires high level of precision in tuning the interaction between qubits. Therefore a higher precision in programming the problem Hamiltonian is an essential feature for a quantum annealer.

*Coherence-* Decoherence can be a major limitation for analog quantum computers since there is no established error correction formalism for fault-tolerant AQC. Therefore, one expects higher success for AQC with better device coherence. However, analog AQC has noise resilient properties which might be lost in a gate-based digital implementation [23, 24].

Each of the above features adds to the hardware complexity of an analog quantum annealer. It would be realistic to say that any design would inevitably compromise some of these elements. A digital approach to AQC, however, has no fundamental limit to achieve the above features since it simulates AQC with single and two-qubit gates. Of course there is a cost in terms of required qubits, which will be discussed next.

## II. METHODS OF DIGITIZATION AND DISCUSSION OF SCALING

The experiments in this paper explore the digitization of adiabatic quantum computing using the first-order Lie-Trotter-Suzuki formula [25, 26]. We now discuss the scaling of the number of gates that this scheme requires to prepare the target state to within fixed error. We restrict our focus to adiabatic evolutions under time-dependent Hamiltonians that are decomposable into $L$ different $k$-local Hamiltonians such that $H(t) = \sum_{\ell=1}^{L} a_\ell(t) H_\ell$ where the $a_\ell(t)$ are time-dependent scalars and the $H_\ell$ are local Hamiltonians having many-body order of at most $k$ [27]. We approximate a continuous time evolution for time $T$ by discretizing time into steps of equal size, $\delta t = T/M$ where $M$ is the number of time steps. In our experiment the digitization of the continuous time evolution is simulated as

$$U_{\text{digital}} = \prod_{m=1}^{M} \prod_{\ell=1}^{L} \exp\left[-i\delta t \, a_\ell(m\delta t) \, H_\ell\right]. \quad \text{(S1)}$$

Assuming the ability to implement arbitrary rotations, evolution under any $k$-local Hamiltonian can be implemented using a number of gates that is at most $\mathcal{O}(k)$. Thus, the gate complexity of this approach is $\mathcal{O}(MLk)$. We now address how $M$ should be chosen to perform a continuous time evolution $U_{\text{continuous}}$ such that $\|U_{\text{continuous}} - U_{\text{digital}}\| \leq \epsilon$. Here $\epsilon$ upper-bounds the largest error that can be induced on any eigenstate of the Hamiltonian. While the strategy is not employed in our experiment, one can derive a significantly tighter bound on the discretization error by making the following substitution to Eq. S1 [28],

$$\delta t \, a_\ell(m\delta t) \to \int_{(m-1)\delta t}^{m\delta t} a_\ell(s) \, ds. \quad \text{(S2)}$$

In [29, 30], it is shown that such an evolution can be simulated with error $\epsilon$ by choosing $M = T^2 a_{\max}^2 L^2 / \epsilon$ where $a_{\max} = \max_{\ell, t} \{a_\ell(t)\}$. The adiabatic theorem [1] dictates that $T$ should be chosen as,

$$T = \mathcal{O}\left(\frac{\max_t \left|\langle 1; t| \frac{dH(t)}{dt} |0; t\rangle\right|}{\gamma^2}\right) = \mathcal{O}\left(\frac{D}{\gamma^2}\right) \quad \text{(S3)}$$

where $\gamma$ is the minimum spectral gap during the adiabatic evolution, $|0; t\rangle$ and $|1; t\rangle$ denote the ground and first excited state at time $t$. Putting these bounds together we find that the total number of gates should scale as

$$\mathcal{O}(MLk) = \mathcal{O}\left(\frac{T^2 a_{\max}^2 L^3 k}{\epsilon}\right) = \mathcal{O}\left(\frac{a_{\max}^2 D^2 L^3 k}{\gamma^4 \epsilon}\right). \quad \text{(S4)}$$

We chose simple first order Trotterization for this experiment only because of experimental limitations. Since, due to substantial overhead in $L$, $k$ and $\gamma^{-1}$, it is unlikely that an approach based on the first order Trotter decomposition will be of practical use. With a future device of larger size and better coherence, we would be able to significantly improve the method of digitization. For instance, a digital simulation scheme based on the truncation of the Taylor series of the time-evolution operator [31] has been shown to exponentially outperform Trotterization in terms of $\epsilon$, scale linearly with $T$ (up to logarithmic factors), which implies a quadratic reduction in $\gamma^{-1}$, and scale much better with the number of terms for real world applications such as the simulation of chemistry [32, 33]. As quantum hardware improves, the implementation of near-optimal schemes such as this becomes increasingly viable.

## III. RESIDUAL ENERGY SCALING

Here, we motivate the residual energy and its scaling with simulated time and number of spins. We find that for large Ising spin chains the residual energy follows a power law, $T^{-\eta}$ with $\eta \geq 0.5$.

The standard picture of AQC describes that the ground state of a Hamiltonian (problem Hamiltonian) encodes the solution of a computational problem. Quantum adiabatic theorem tells us how slowly we should drive the system to reach to this target state with high probability. Therefore a measure of success for computation is the overlap between the system state at the end of the evolution and the ground state of the problem Hamiltonian. Such a measure might be unnecessary for optimization problems as most of the time reaching a good local minimum could be satisfactory instead of the global minimum of the ground state. Therefore a relevant measure would be the residual energy above the ground state, the smaller the better.

The Kibble-Zurek mechanism and the Landau-Zener theorem are consistent approaches to estimate the residual energy for a many-body system that slowly passes through a phase transition [34–37]. In the experiment, Fig. 3 in the main article, a transverse field drives a chain of spins with Ising interaction through phase transition at different speeds. The KZ mechanism explains that as a system goes faster through the critical point, there would be less time for spins to communicate in order to find the ground states. That translates into an incomplete formation of the ground state and the emergence of kinks after the phase transition. The density of kinks is monotonically related to the energy of the final exited state. Here we follow the line of argument in Ref. [35] to find the scaling of residual energy at fast and intermediate speeds. Note that the scaling theory describes dynamics in thermodynamical limit of a large number of spins. However, we see a correspondence between our few-qubit experiment and the scaling analysis presented below.

For fast quench, a system that starts in the superposition of all energy levels (ground state of uniform transverse Hamiltonian) has insufficient time to adjust to lower energy states. Therefore the system stays close to its initial energy distribution with little dependence on the time $T$. This appears as a plateau in the residual energy plot. For longer evolution times $T$, we consider a two-level system approximation and apply the Landau-Zener formula that gives the probability of excitement into the first exited state as $P = e^{-\alpha \Delta^2 T}$. For an $N-$spin system of the experiment, with a uniform Ising Hamiltonian, $\Delta$, the minimum gap, scales as $\frac{1}{N}$. For a fixed time $T$ and likelihood $p^*$, we find the longest defect-free chain as $N^* = |\frac{\alpha}{\ln p^*}|^{\frac{1}{2}} T^{1/2}$. Therefore, to first order the kink density and residual energy scale as $\frac{1}{N^*} \propto T^{-1/2}$. Although our residual energy experiment is small-scale, the two phases of plateau for short times, and a transition to a power law $T^{-\eta}$ with $\eta > 0.5$ are visible.

## IV. PAIRWISE INTERACTION IN A NINE QUBIT SYSTEM

The nine qubit chain is placed in a configuration with alternating frequencies for idling, to minimize parasitic interactions from nearby qubits. Adjacent qubits are detuned by typically 1 GHz, and next-nearest qubits are detuned by 0.1 GHz. The idling configuration is shown in Fig. S1a. Figure S1b shows the implementation of entangling, for example between qubits $Q_1$ and $Q_0$, we move $Q_0$ to a higher frequency, and let $Q_1$ undergo an adiabatic trajectory which is tuned to bring about a conditional phase shift while minimizing state leakage [38, 39]. We apply decoupling pulses to $Q_2$ during this interaction. Other qubits undergo entangling gates at the same time. After this interaction, the qubit frequencies are returned to the idling positions.

## V. CONSTRUCTING INTERACTION

At the core of the multibody interactions is the $CZ_\phi$ entangling gate. One qubit is held at a steady frequency while the other undergoes an adiabatic trajectory which sweeps $|02\rangle$ close to $|11\rangle$ [38]. By varying the amplitude of this trajectory we can tune the conditional phase [40]. After this interaction we null the single qubit phases, arising from the single qubit frequency detunings. We find that by careful calibrations, we can achieve the desired conditional phase, and null the single qubit phases to within 0.05 rads, see Fig. S2a-b. Other qubits are decoupled from this interaction with $\pi$ rotations, see below.

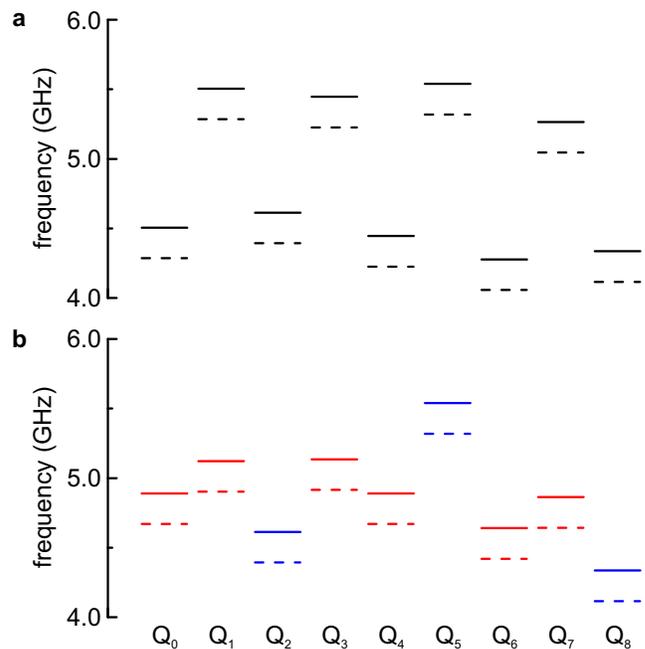

FIG. S1. **Frequency configuration for idling and interacting.** (a) Idling configuration, showing an alternating frequency pattern designed to minimize interaction. (b) Configuration where qubits interact. Black: idling qubits. Red: adjacent pairs of qubits are performing a $CZ_\phi$ entangling gate. Blue: qubits which are decoupled using $\pi$ pulses. This configuration corresponds to the dashed line in Fig. S3a.

The tunable phase is limited between $\phi \sim 0.5$ to $\sim 4.5$. Below this range interactions with other qubits complicate implementation, and above this range population leakage into higher-energy levels becomes significant. In order to construct a tunable gate over the full range, we choose one of three implementations based on the desired phase for $\sigma_z \otimes \sigma_z$:

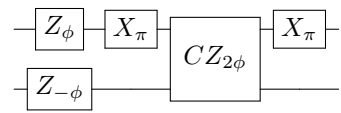

For $\phi > 0.25$:

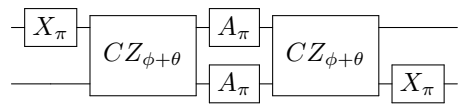

For $-0.25 < \phi < 0.25$:

with $A = Y$ and $\theta = \pi$.

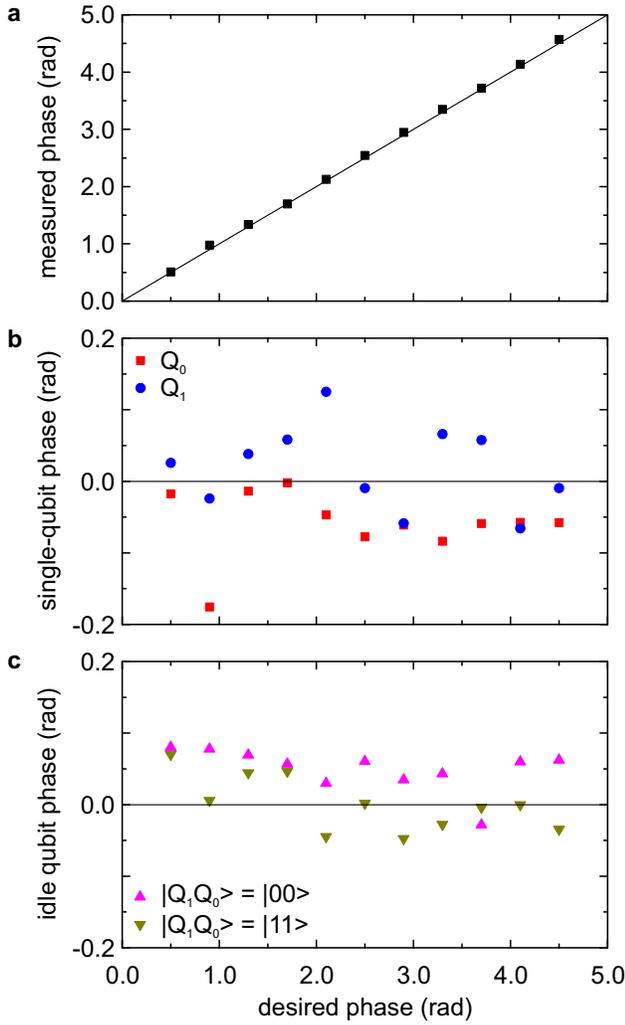

FIG. S2. **Phases for the CZ$_\phi$ gate.** (a) Measured vs. desired controllable phase $\phi$. (b) Residual single-qubit phases. (c) Residual phase on the idling qubit Q$_2$ for $|Q_1Q_0\rangle = |00\rangle$ and $|Q_1Q_0\rangle = |11\rangle$.

And for $\phi < -0.25$:

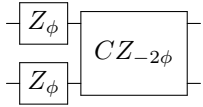

And for $|\phi| > 2.25$ we either apply the quantum circuit with two entangling gates with $A = X$ and $\theta = 0$, or add or subtract $2\pi$ until it is in range. Implementing the unitary $U = \exp(-iJ_{zz}\Delta t)$ is then done by setting $\phi = 2J_{zz}\Delta t$. The above identities ensure we can implement any strength of $\sigma_z \otimes \sigma_z$.

## VI. DECOUPLING FROM THE ENVIRONMENT AND PARASITIC INTERACTIONS

Our qubits have dephasing, dominated by correlated processes, and are susceptible to parasitic interactions with other qubits [41]. To reduce these effects, we include decoupling $\pi$ pulses in three locations in the algorithm: I) Around the $\sigma_z\sigma_z$ and $\sigma_x\sigma_x$ interaction. At the start of a $\sigma_z\sigma_z$ or $\sigma_x\sigma_x$ interaction we apply an X$_\pi$ rotation on both qubits; at the end we apply an X$_{-\pi}$ rotation. This maintains the unitary but decreases the effects of qubit dephasing. II) Idling qubits are decoupled from the environment by applying two X$_\pi$ pulses, centered at $\tau/4$ and at $3\tau/4$, with $\tau$ being the idling time. III) Qubits which are adjacent to a qubit undergoing a controlled-phase frequency trajectory need to be decoupled from this interaction; in contrast to the idling case, we now apply closely spaced sequential X$_\pi$ and X$_{-\pi}$ rotations during the frequency trajectory of the other qubit, to null the parasitic interaction. We find mean phase errors from residual parasitic interaction to be around 0.05 rad, which is equivalent to a gate error of $1 - \cos^2(\frac{\delta\phi}{2}) = 6 \cdot 10^{-4}$, see Fig. S2c. With decoupling pulses the errors are dominated by intrinsic gate errors.

## VII. PULSE SEQUENCES

The evolution is digitized using the Trotter expansion [25, 26]. In essence the evolution is divided into many small steps in time, $U = \mathcal{T}\exp(-i\int H(t)dt) \simeq \exp(-iH(t_1)\Delta t)\exp(-iH(t_2)\Delta t)...$, where each $H(t_n)$ is comprised of terms which sum up to $H(t_n) = \sum_\ell H_\ell(t_n)$, implemented using a construction of quantum logic gates. Local fields come from single qubit gates, and the full range of $\sigma_z\sigma_z$ and $\sigma_x\sigma_x$ couplings come from one or two CZ$_\phi$ gates in a combination with single qubit gates.

The pulse sequences are shown in Fig. S3 for the scaling experiment with nine qubits (Fig. 3 in the main article), and for the random stoquastic and non-stoquastic problems with six to nine qubits (Fig. 5 in the main article). Slow, rectangular-like pulses are frequency detunings, and rapidly oscillating waveforms denote microwave pulses. Numbers in the figure correspond to the following:

1. Initial state preparation: $|+\rangle^{\otimes N}$, with $N = 9$ qubits.

2. First Trotter step

3. Second Trotter step

TABLE S1. Gate counts for pulse sequences in Fig. S3. We count idles as any duration of 10 ns or longer. Long idles are counted as a single idle, even though the relevant approach for estimating total process fidelities is by splitting idles in terms of durations of the microwave gates [superconducting circuits]. The gate counts are for the full algorithm, all Trotter steps as well as initialization.

|  | a | b | c | d | e |
|---|---|---|---|---|---|
| entangling CZ$_\phi$ gates | 16 | 29 | 52 | 26 | 18 |
| single qubit gates | 263 | 550 | 1059 | 486 | 326 |
| - microwave $\pi$ and $\pi/2$ | 135 | 292 | 598 | 282 | 173 |
| - idle | 78 | 178 | 331 | 142 | 103 |
| - virtual phase | 50 | 80 | 130 | 62 | 50 |

4. $\sigma_z\sigma_z$ interaction: consisting of decoupling pulses, $Q_0$ being moved to an interaction frequency, and $Q_1$ performing the trajectory.

5. Environmental decoupling pulses: $\pi$ pulses around $\sigma_z\sigma_z$, but such that the ideal unitary is unchanged

6. Decoupling pulses to reduce parasitic interactions with idling qubits

7. Environmental decoupling pulses

8. Implementation of $B_x$. The decrease in amplitude with Trotter step reflects the decrease in $B_x$ following the annealing schedule.

9. $\sigma_z\sigma_z$ interaction for an angle which requires 2 CZ$_\phi$ gates.

10. First Trotter step of the non-stoquastic problem evolution, showing both $\sigma_z\sigma_z$ and $\sigma_x\sigma_x$ interaction. Note that other qubits have to wait if a pair has an interaction strength which requires 2 CZ$_\phi$ gates.

11. $\sigma_z\sigma_z$ interaction with large and small angles

12. $\sigma_x\sigma_x$ interaction, showing the $\pi/2$ pulses for basis rotation.

13. Notice how $\sigma_x\sigma_x$ interaction in the second Trotter step is now done with only a single CZ$_\phi$ gate. The coupling strength is linearly turned on and phases increase, allowing the interaction to be implemented with a single entangling gate.

14. As a result the Trotter steps 2-5 are shorter than the first.

15. Qubit frequency configuration shown in Fig. S1b.

A gate count of the sequences is provided in Table S1.

## VIII. SIMULATION PARAMETERS

An overview of the number of Trotter steps, simulated times, coupling and field strengths for the performed experiments is shown in Table S2. For the single instances shown in the main article see Tables S4-S3.

## IX. DIGITAL EVOLUTION INTO GHZ STATE: IMAGINARY PARTS AND IDEAL ADIABATIC EVOLUTION

The real and imaginary parts, as well as the ideal adiabatic evolution and target state for the experiment in the main article in Fig. 2 at $s = 1.0$ are shown in Fig. S4. The fidelity of the ideal adiabatic evolution with respect to the target state is 0.92. The fidelity of the experimental data with respect to the ideal adiabatic evolution is 0.60.

## X. KINK LIKELIHOOD FOR TWO TO NINE QUBIT CONFIGURATIONS

The kink likelihood for configurations with two to nine qubits is shown in Fig. S5. For two qubits, only a single kink is possible ($|01\rangle$ or $|10\rangle$), and initially no kink or a single kink are equally likely. When increasing the simulation time the kink likelihood decreases, and the likelihood of no kinks increases. This picture is repeated for all systems. For the seven to nine qubit systems, around $|J|T = 2$, the likelihood of kinks increases again. The experimental data closely follow the ideal digital evolution (dashed).

The differences in residual energies are plotted in Fig. S6, for three, six and nine qubits. The increase in difference for the nine qubit system near $|J|T = 3$ (dashed and dotted blue lines) is due to digitization error, as the experiment follows the ideal digital evolution to within a difference of $2|J|$ (solid blue line).

## XI. SPIN PARITY CORRELATION

Spin parity correlations for the five qubit antiferromagnetic experiment are displayed in Fig. S7 as a function of distance $d$ and magnetic field. The measured parity correlations (left) reflect the anti-ferromagnetic nature: the correlation is negative for odd distances and positive otherwise. The correlations follow the theory predictions (right) for either direction.

## XII. GATE CALIBRATIONS

Variable single qubit microwave rotations are calibrated by inferring the rotation angle from measurements of the probability with amplitude (Fig. S8a). The tunable CZ$_\phi$ gate is calibrated by placing the static qubit in an equator state, and placing the other qubit in either $|0\rangle$ or $|1\rangle$, and varying the amplitude of the trajectory. By performing quantum state tomography on the static qubit the tunable phase becomes apparent, see Fig. S8b.

## XIII. COMPARISON OF FIDELITIES BETWEEN EXPERIMENT, IDEAL DIGITAL EVOLUTION, IDEAL CONTINUOUS EVOLUTION, AND TARGET STATE RESULTS, AS WELL AS UNIFORM RANDOM VALUES

Here, we list comparisons between experiment, ideal digital evolution, ideal continuous evolution as well as the target state results, expressed as fidelities. We also show comparisons with uniformly chosen random probabilities as a baseline sanity check. A complete overview is shown in Table S9.

We also show the histograms of all fidelities in Fig. S9.

Fidelities for the example instances of Fig. 5 in the main article are in Table S10.

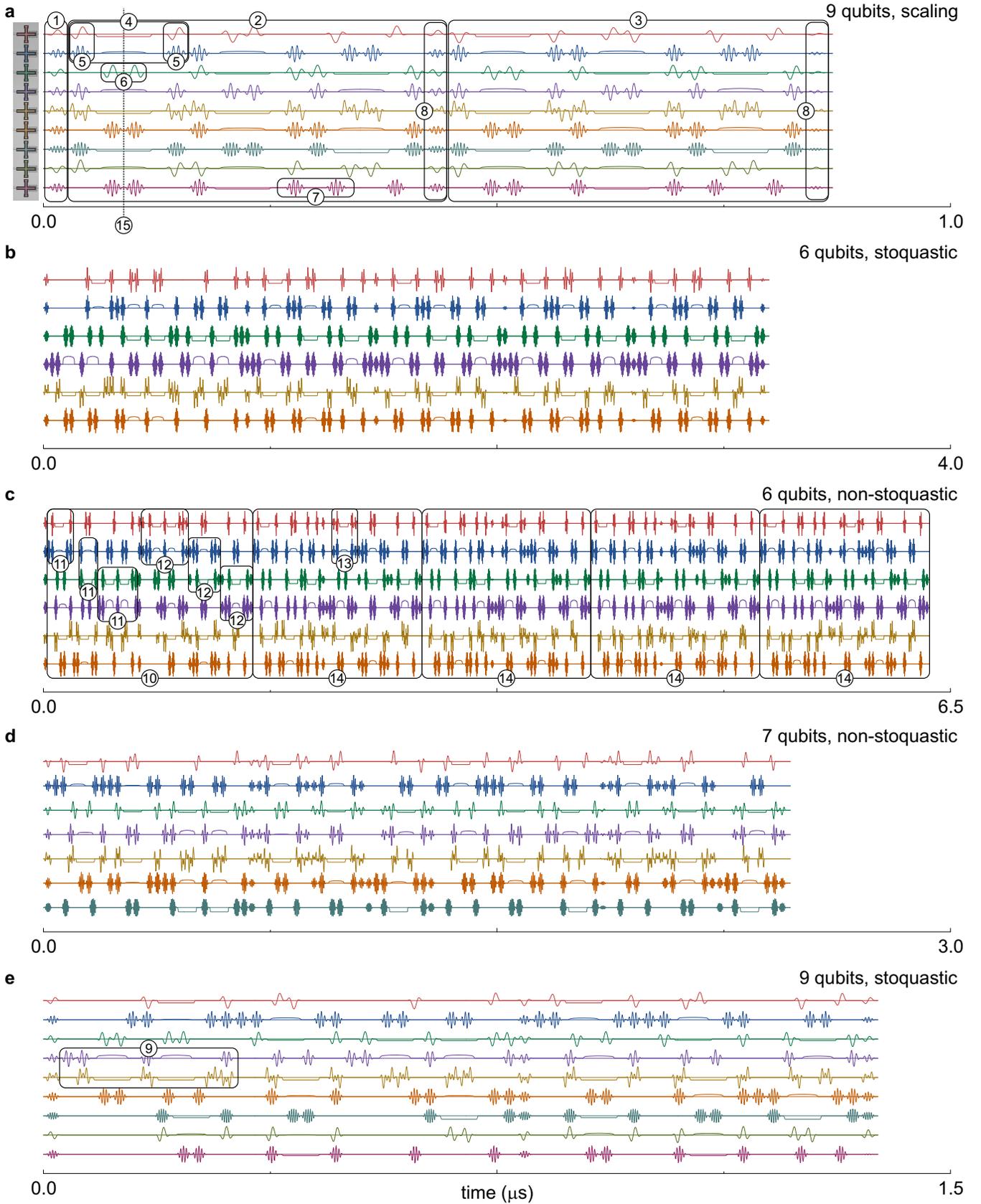

FIG. S3. **Pulse sequences.** (a) Nine qubit scaling experiment at $T = 1$ from Fig. 3 in the main article. (b) Six qubit stoquastic problem from Fig. 5b in the main article. (c) Six qubit non-stoquastic problem from Fig. 5e in the main article. (d) Seven qubit non-stoquastic problem. (e) Nine qubit stoquastic problem. The numbers in the figures are explained in the main text.

TABLE S2. Simulation parameters for the experiments. Random problem denotes both the stoquastic and non-stoquastic one.

| experiment | coupling | local field | Trotter steps | simulated time |
|---|---|---|---|---|
| ferromagnetic chain | $J_{zz} = 2$ | None | 5 | $T = 3$ |
| scaling (2-6 qubits) | $J_{zz} = 2$ | None | 3 | $T = 0...1.5$ |
| scaling (7-9 qubits) | $J_{zz} = 2$ | None | 2 | $T = 0...1.5$ |
| AF chain w. local field | $J_{zz} = -1.25$ | middle qubit: $B_z = -3...3$ | 4 | $T = 2.5$ |
| random problem, 3 qubits | $-2... - 0.5$ or $0.5...2$ | $-2...2$ | 5 | $T = 3$ |
| random problem, 6 qubits | $-2... - 0.5$ or $0.5...2$ | $-2...2$ | 5 | $T = 3$ |
| random problem, 7 qubits | $-2... - 0.5$ or $0.5...2$ | $-2...2$ | 2 | $T = 1$ |
| random problem, 8 qubits | $-2... - 0.5$ or $0.5...2$ | $-2...2$ | 2 | $T = 1$ |
| random problem, 9 qubits | $-2... - 0.5$ or $0.5...2$ | $-2...2$ | 2 | $T = 1$ |

TABLE S3. Nine qubit stoquastic problem instance.

| | $Q_0$ | $Q_1$ | $Q_2$ | $Q_3$ | $Q_4$ | $Q_5$ | $Q_6$ | $Q_7$ | $Q_8$ |
|---|---|---|---|---|---|---|---|---|---|
| $B_x$ | 1.437 | 0.749 | 0.912 | 1.153 | 1.523 | 1.670 | 1.621 | 1.930 | -0.899 |
| $B_z$ | -0.559 | -1.078 | -1.822 | -0.407 | 0.652 | 1.675 | 1.362 | 0.302 | -0.187 |
| $J_{zz}$ | | -0.781 | -1.672 | 0.520 | 0.635 | 0.812 | -0.816 | 1.162 | 0.639 |

TABLE S4. Three qubit stoquastic problem instance.

|        | $Q_0$  | $Q_1$  | $Q_2$  |
|--------|--------|--------|--------|
| $B_x$  | -0.159 | 1.22   | -1.93  |
| $B_z$  | -1.29  | -1.45  | -0.772 |
| $J_{zz}$ |     | -1.09  | 1.16   |

TABLE S5. Three qubit non-stoquastic problem instance.

|        | $Q_0$  | $Q_1$  | $Q_2$  |
|--------|--------|--------|--------|
| $B_x$  | -1.18  | -1.71  | 1.02   |
| $B_z$  | -0.875 | 0.781  | -0.428 |
| $J_{xx}$ |     | -0.841 | 1.02   |
| $J_{zz}$ |     | -0.757 | 1.32   |

TABLE S6. Six qubit stoquastic problem instance.

|        | $Q_0$  | $Q_1$  | $Q_2$  | $Q_3$  | $Q_4$  | $Q_5$  |
|--------|--------|--------|--------|--------|--------|--------|
| $B_x$  | 0.155  | -1.238 | 1.789  | 0.899  | -1.501 | -1.309 |
| $B_z$  | 0.468  | -1.577 | -1.183 | -0.665 | -0.928 | -1.265 |
| $J_{zz}$ |     | 1.476  | -0.740 | -0.765 | -0.535 | -0.966 |

TABLE S7. Six qubit non-stoquastic problem instance.

|        | $Q_0$  | $Q_1$  | $Q_2$  | $Q_3$  | $Q_4$  | $Q_5$  |
|--------|--------|--------|--------|--------|--------|--------|
| $B_x$  | -0.255 | 0.606  | -1.735 | 0.732  | 1.586  | -0.305 |
| $B_z$  | -1.672 | -1.282 | -1.532 | -1.433 | 1.282  | -1.765 |
| $J_{xx}$ |     | 0.577  | -1.954 | -1.616 | -1.517 | -1.896 |
| $J_{zz}$ |     | -1.491 | 1.349  | 0.628  | 1.287  | 1.919  |

TABLE S8. Seven qubit non-stoquastic problem instance.

|        | $Q_0$  | $Q_1$  | $Q_2$  | $Q_3$  | $Q_4$  | $Q_5$  | $Q_6$  |
|--------|--------|--------|--------|--------|--------|--------|--------|
| $B_x$  | -1.335 | 0.760  | -1.261 | -0.221 | -0.892 | -1.321 | 0.133  |
| $B_z$  | -1.026 | -1.896 | 0.116  | -0.619 | -0.493 | -1.316 | -1.872 |
| $J_{xx}$ |     | 1.891  | 1.517  | 1.568  | 0.748  | 1.419  | -0.839 |
| $J_{zz}$ |     | -1.455 | -0.588 | -0.582 | 1.223  | -0.635 | 0.614  |

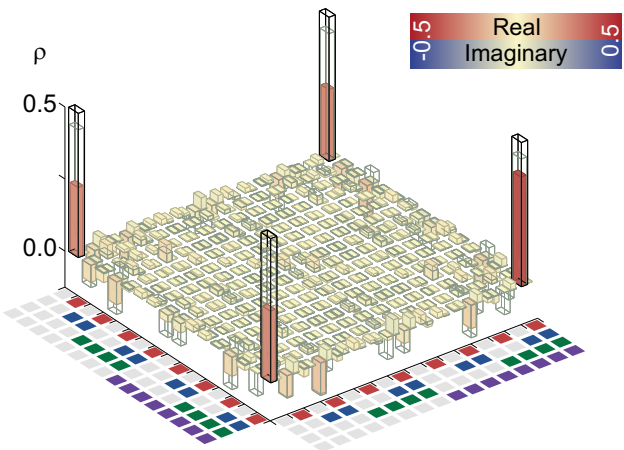

FIG. S4. **Digital evolution into GHZ state: real, imaginary parts and ideal continuous time evolution.** Experimental data (colour), target state (black) and ideal continuous time evolution (gray) at $s = 1.0$. The leftmost red bars indicate the real part, the adjacent blue bars indicate the imaginary part. $\mathrm{Im}(\rho) < 0.05$.

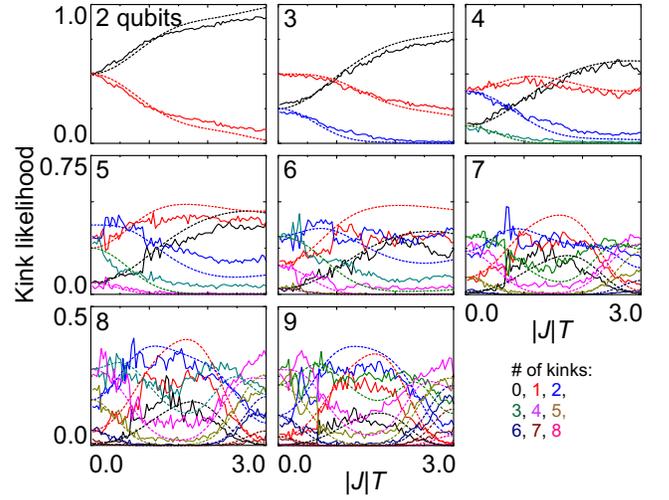

FIG. S5. **Kink likelihood for two to nine qubit configurations.** Errors in ferromagnetic chains ($J = 2$) in configurations with two to nine qubits. Kink likelihood versus scaled time $|J|T$. Solid lines: experiment. Dashed lines: ideal digital evolution.

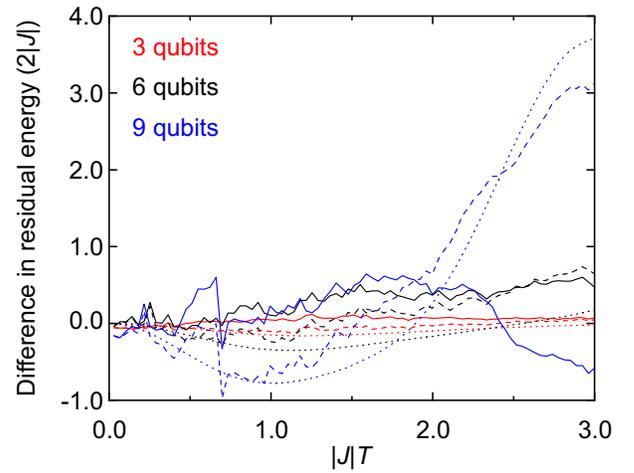

FIG. S6. **Differences between residual energy for the experimental, ideal digital, and ideal continuous evolutions.** Residual energy difference for three, six, and nine qubits. Shown are the differences between the experiment and ideal digital evolution (solid lines), experiment and ideal continuous evolution (dashed), and between the ideal digital and continuous evolutions (dotted).

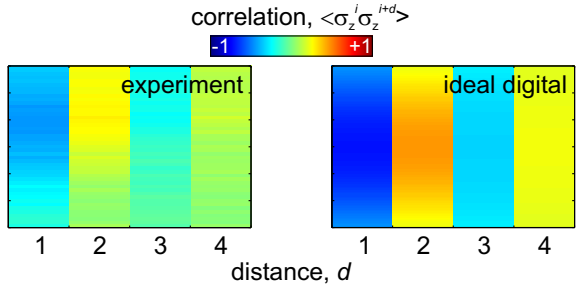

FIG. S7. **Spin parity correlations in the antiferromagnetic state during lifting of degeneracy.** Left: Experimental data. Right: Ideal digitized evolution.

TABLE S9. Mean fidelities between the experimental data, the ideal digital evolution, ideal continuous evolution (ideal cont.) and target state. As a baseline sanity check, we also show comparisons with randomly generated data. The standard deviations from the mean are given.

| data | ideal digital evolution | ideal continuous evolution | target state |
|---|---|---|---|
| 3 qubits, stoquastic | | | |
| experiment | $0.706 \pm 0.007$ | $0.48 \pm 0.01$ | $0.36 \pm 0.02$ |
| ideal digital | | $0.67 \pm 0.01$ | $0.46 \pm 0.02$ |
| ideal cont. | | | $0.61 \pm 0.03$ |
| 3 qubits, non-stoquastic | | | |
| experiment | $0.36 \pm 0.01$ | $0.220 \pm 0.009$ | $0.185 \pm 0.009$ |
| ideal digital | | $0.49 \pm 0.02$ | $0.29 \pm 0.02$ |
| ideal cont. | | | $0.53 \pm 0.03$ |
| 6 qubits, stoquastic | | | |
| experiment | $0.714 \pm 0.006$ | $0.523 \pm 0.008$ | $0.296 \pm 0.007$ |
| random | $0.496 \pm 0.007$ | $0.340 \pm 0.007$ | $0.168 \pm 0.005$ |
| ideal digital | | $0.73 \pm 0.01$ | $0.43 \pm 0.01$ |
| ideal cont. | | | $0.59 \pm 0.01$ |
| 6 qubits, non-stoquastic | | | |
| experiment | $0.739 \pm 0.004$ | $0.522 \pm 0.008$ | $0.380 \pm 0.009$ |
| random | $0.669 \pm 0.004$ | $0.470 \pm 0.007$ | $0.335 \pm 0.008$ |
| ideal digital | | $0.526 \pm 0.009$ | $0.350 \pm 0.009$ |
| ideal cont. | | | $0.62 \pm 0.01$ |
| 7 qubits, stoquastic | | | |
| experiment | $0.645 \pm 0.006$ | $0.607 \pm 0.006$ | $0.215 \pm 0.009$ |
| random | $0.543 \pm 0.006$ | $0.534 \pm 0.006$ | $0.133 \pm 0.006$ |
| ideal digital | | $0.883 \pm 0.004$ | $0.332 \pm 0.009$ |
| ideal cont. | | | $0.281 \pm 0.009$ |
| 7 qubits, non-stoquastic | | | |
| experiment | $0.632 \pm 0.006$ | $0.566 \pm 0.006$ | $0.311 \pm 0.009$ |
| random | $0.607 \pm 0.005$ | $0.553 \pm 0.006$ | $0.277 \pm 0.008$ |
| ideal digital | | $0.812 \pm 0.006$ | $0.34 \pm 0.01$ |
| ideal cont. | | | $0.36 \pm 0.01$ |
| 8 qubits, stoquastic | | | |
| experiment | $0.606 \pm 0.006$ | $0.570 \pm 0.006$ | $0.164 \pm 0.007$ |
| random | $0.513 \pm 0.006$ | $0.509 \pm 0.006$ | $0.091 \pm 0.004$ |
| ideal digital | | $0.873 \pm 0.004$ | $0.274 \pm 0.008$ |
| ideal cont. | | | $0.225 \pm 0.007$ |
| 8 qubits, non-stoquastic | | | |
| experiment | $0.572 \pm 0.005$ | $0.499 \pm 0.006$ | $0.245 \pm 0.008$ |
| random | $0.585 \pm 0.005$ | $0.517 \pm 0.005$ | $0.238 \pm 0.007$ |
| ideal digital | | $0.775 \pm 0.006$ | $0.292 \pm 0.009$ |
| ideal cont. | | | $0.32 \pm 0.01$ |
| 9 qubits, stoquastic | | | |
| experiment | $0.583 \pm 0.007$ | $0.551 \pm 0.006$ | $0.122 \pm 0.006$ |
| random | $0.496 \pm 0.006$ | $0.481 \pm 0.006$ | $0.074 \pm 0.004$ |
| ideal digital | | $0.862 \pm 0.004$ | $0.228 \pm 0.007$ |
| ideal cont. | | | $0.184 \pm 0.007$ |
| 9 qubits, non-stoquastic | | | |
| experiment | $0.587 \pm 0.006$ | $0.507 \pm 0.006$ | $0.236 \pm 0.008$ |
| random | $0.570 \pm 0.004$ | $0.495 \pm 0.005$ | $0.214 \pm 0.008$ |
| ideal digital | | $0.747 \pm 0.006$ | $0.248 \pm 0.008$ |
| ideal cont. | | | $0.27 \pm 0.01$ |

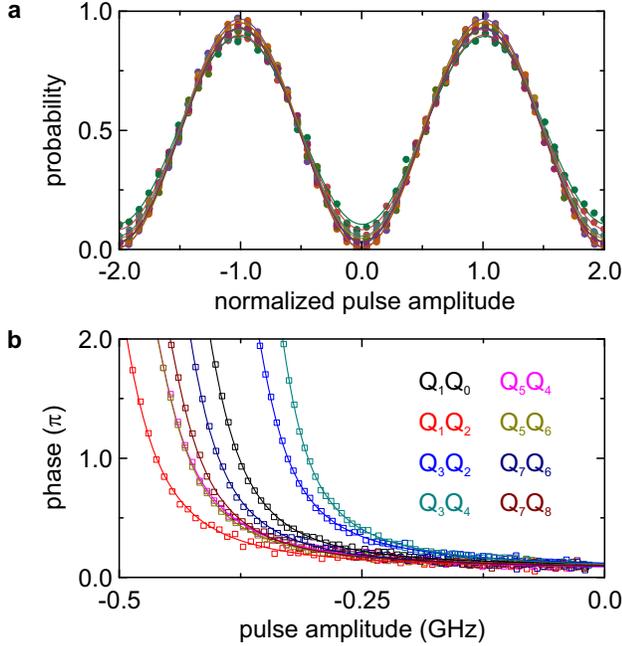

FIG. S8. **Gate calibrations.** (a) Microwave rotation pulse calibrations for all nine qubits. Pulse amplitude is normalized to the amplitude of a $\pi$-pulse. Solid lines are fits to the data. Colours are linked to Fig. 1b in the main article. (b) Controllable phase $\phi$ of the $CZ_\phi$ gate versus qubit detuning for all eight adjacent pairs. The difference between the curves is due to the qubits having different setpoints in frequency. Solid lines are fits to the data.

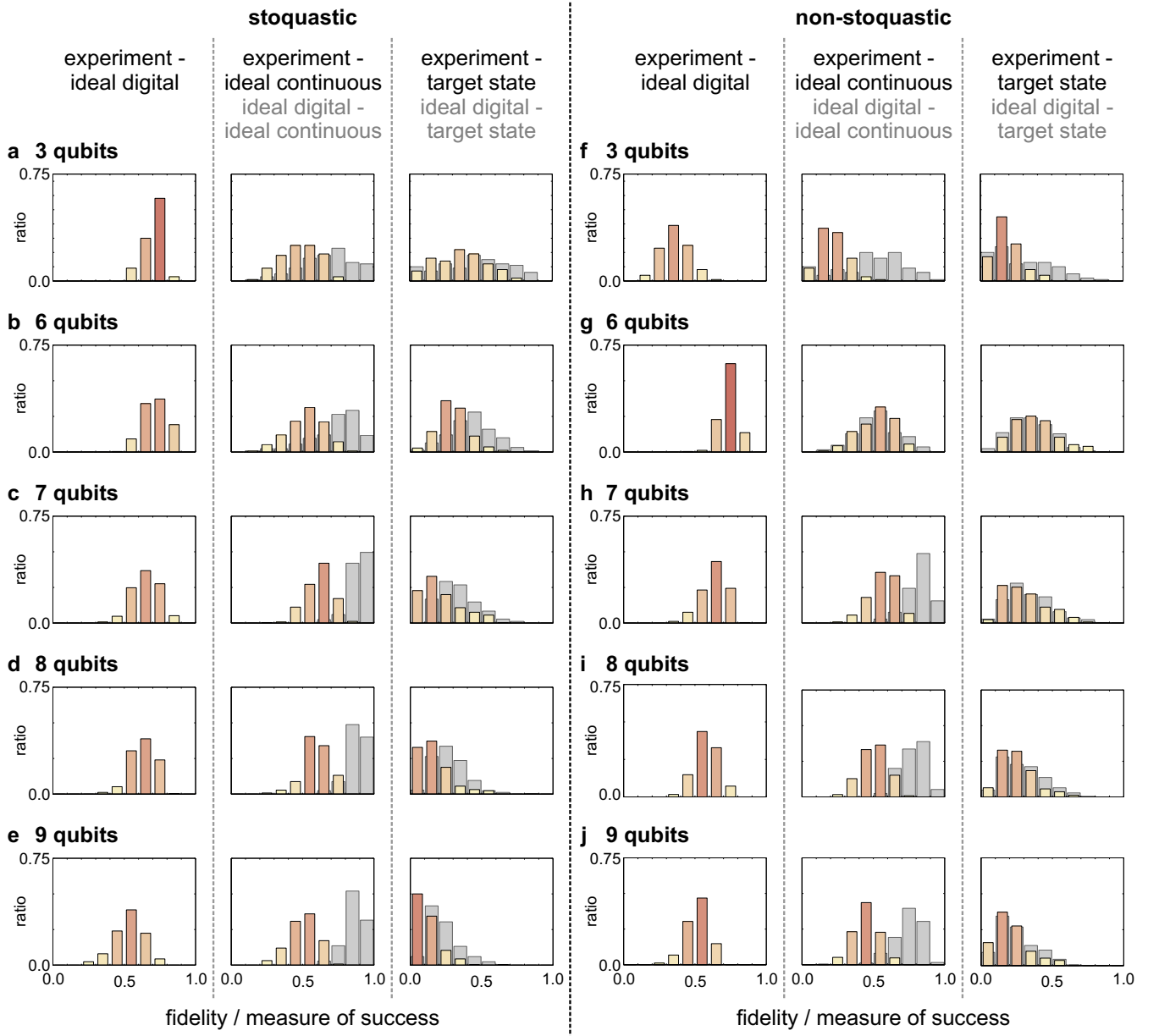

FIG. S9. **Fidelity overview for the random stoquastic and non-stoquastic problems.** Here, we show normalized histograms of fidelities and measures of success for all experiments for both stoquastic and non-stoquastic problems. For each case we plot three histograms: On the left, we plot the fidelity of the experimental results with respect to the ideal digital evolution. In the middle, in colour we plot the fidelity of the experimental results with respect to the ideal, continuous time evolution for finite time. In the same figure, we plot in gray the fidelity of the ideal digital results with respect to the continuous, finite time evolution. On the right, we plot in colour the fidelity of the experimental results with respect to the ideal adiabatic evolution, and in gray the fidelity of the ideal digital results with respect to the ideal adiabatic evolution. (a, b, c, d, e) Results for random stoquastic problems with three, six, seven, eight and nine qubits. (f, g, h, i, j) Results for random non-stoquastic problems with three, six, seven, eight and nine qubits.

TABLE S10. Fidelities and measures of success for the example instances in Fig. 5 between the experimental data, the ideal digital evolution, ideal continuous evolution (ideal cont.) and target state.

| data | ideal digital evolution | ideal continuous evolution | target state |
|------|------------------------|----------------------------|--------------|
| 3 qubits, stoquastic | | | |
| experiment | 0.70 | 0.68 | 0.63 |
| ideal digital | | 0.96 | 0.90 |
| ideal cont. | | | 0.92 |
| 3 qubits, non-stoquastic | | | |
| experiment | 0.48 | 0.43 | 0.42 |
| ideal digital | | 0.82 | 0.60 |
| ideal cont. | | | 0.59 |
| 6 qubits, stoquastic | | | |
| experiment | 0.78 | 0.69 | 0.65 |
| ideal digital | | 0.90 | 0.61 |
| ideal cont. | | | 0.62 |
| 6 qubits, non-stoquastic | | | |
| experiment | 0.76 | 0.72 | 0.42 |
| ideal digital | | 0.65 | 0.38 |
| ideal cont. | | | 0.52 |
| 9 qubits, stoquastic | | | |
| experiment | 0.66 | 0.60 | 0.63 |
| ideal digital | | 0.83 | 0.64 |
| ideal cont. | | | 0.77 |
| 7 qubits, non-stoquastic | | | |
| experiment | 0.71 | 0.76 | 0.71 |
| ideal digital | | 0.77 | 0.53 |
| ideal cont. | | | 0.69 |